\documentclass[submission, Phys]{SciPost}
\usepackage[T1]{fontenc}
\usepackage{bm}
\usepackage{amsmath}
\usepackage{amsfonts}
\usepackage{mathtools}
\usepackage{amssymb}
\usepackage[dvipsnames]{xcolor}
\usepackage{hyperref}
\usepackage{graphicx}
\usepackage[space]{grffile}
\usepackage{txfonts}

\usepackage[normalem]{ulem}

\begin{document}
\begin{center}{\Large \textbf{
Quantum Thermalization beyond Non-Integrability and Quantum Scars in a Multispecies Bose-Josephson Junction
}}\end{center}

\begin{center}
Francesco Di Menna\textsuperscript{1,2*},
Sergio Ciuchi\textsuperscript{1},
Simone Paganelli\textsuperscript{1,2}
\end{center}

\begin{center}
{\bf 1} Dipartimento di Scienze Fisiche e Chimiche, Universit\'a dell’Aquila, Coppito-L’Aquila, Italy
\\
{\bf 2} INFN, Laboratori Nazionali del Gran Sasso, Via G. Acitelli 22, 67100 Assergi (AQ), Italy
\\
* francesco.dimenna@graduate.univaq.it
\end{center}

\begin{center}
\today
\end{center}

    
    

\section*{Abstract}
{\bf
The modern framework for quantum thermalization is grounded in the Eigenstate Thermalization Hypothesis (ETH), in which non-integrability and chaos are historically assumed as prerequisites. 
This work investigates this relationship  in a three-species Bose-Josephson Junction (BJJ) with mutual interactions, experimentally achievable in current ultracold-atom platforms.
After a thorough characterization of quantum chaos in this system, we examine the occurrence of thermal behavior expected when ETH holds. We identify three distinct regimes: chaotic, integrable, and separable. Remarkably, quantum thermalization occurs in both the chaotic and integrable regimes, while it breaks down in and near the separable limit — supporting that non-integrability is not a necessary condition for thermalization. Furthermore, since the system exhibits collective phenomena in the semiclassical limit, we identify ergodicity breaking phenomena such as athermal states in the chaotic regime classifiable as quantum scars, which show no signs of thermalization, consistently with a weak form of ETH. 
}

\vspace{10pt}
\noindent\rule{\textwidth}{1pt}
\tableofcontents\thispagestyle{fancy}
\noindent\rule{\textwidth}{1pt}
\vspace{10pt}

\section{Introduction}

The study of thermalization in classical and quantum many-body systems has been one of the most debated fields in recent years. It has long been commonly believed that the existence of a large number of integrals of motion prevented the system from being ergodic, since it was thought that invariant manifolds in phase space acted as obstacles to ergodicity and therefore as the main impediment to thermal behaviour. However, it is now clear that in classical physics, non-integrability and chaos in particular are neither necessary nor sufficient conditions for ergodicity and thermalization to occur on fast time scales \cite{BALDOVIN20251}.\\
In the quantum counterpart, the situation remains a topic of ongoing debate, centered on the intimate link between quantum chaos and the processes of thermalization. In particular, the non-integrability hypothesis is implicitly contained in the Eigenstate Thermalization Hypothesis (ETH) \cite{srednicki1994chaos,deutsch2018eigenstate,polkovnikov2011colloquium,garrison2018does}, now considered one of the modern formulations of quantum thermalization, according to which the reduced density matrix for a subsystem corresponding to any of the excited eigenstates of the system's Hamiltonian is thermal. 
However, even though ETH has been tested in many interacting quantum systems, rigorous proof of whether or not ETH is a necessary condition for thermalization is still lacking \cite{sinha2024classical}. \\
In addition to the emergence of ergodicity and thermalization in a quantum system, understanding their deviation has also become an area of intense research, attributed to phenomena of ergodicity breaking. However, contrary to common assumptions, even highly complex interacting quantum systems do not necessarily exhibit ergodic temporal evolution. A prominent example is the dynamics of an interacting system with disorder, which can display a localized many-body phase (MBL) that strongly violates ergodicity \cite{abanin2019colloquium}. This phenomenon has been experimentally observed in systems with correlated disorder, such as a quasi-periodic potential \cite{schreiber2015observation}. However, there is another important class of phenomena related to the breakdown of ergodicity, which concerns the violation of the strong form of ETH. The latter states that \textit{all} excited eigenstates are typical in thermodynamical sense \cite{BALDOVIN20251}. However, it has been discovered that certain specific initial states resist to thermalization and exhibit long-time coherent oscillations, in contrast to most other initial states which, within the same energy regime, evolve ergodically over time \cite{sinha2024classical,desaules2022extensive}. This phenomenon of breaking thermalization is called quantum scarring 
\cite{heller1984bound}. In any case, despite being strongly non-thermal, scar states represent an infinitesimal fraction of Hilbert space and are immersed in a much larger sea of proper thermal states, as the volume of phase space affected by scar vanishes in the semiclassical limit \cite{serbyn2021quantum}. For this reason, we can speak about a weak form of ETH.\\ 
Furthermore, in recent years, the debate regarding the relationship between thermalization and chaos has intensified, and researchers are seeking the possibility of finding quantum thermalized behaviour even in the absence of non-integrability \cite{BALDOVIN20251,cattaneo2025thermalization,alba2015eigenstate,ikeda2013finite,trotzky2012probing,EEscalingfree,storms2014entanglement,lydzba2024normal,PhysRevLett.116.030401}.\\

An ideal experimental platform for studying the properties of isolated quantum systems is provided by dilute
ultracold atomic gases \cite{pitaevskii2016bose}. The development of optical trapping techniques has enabled the investigation of phenomena
specific to spinor condensates, since the spin degrees of freedom are no longer frozen, as is the case in magnetic traps
\cite{mele2012spin}. In this framework, spinor condensates can be viewed as multicomponent condensates composed of different bosonic
species, each corresponding to a distinct projection of the spin operator along a given quantization axis \cite{PhysRevA.99.063627}.
These advances have stimulated extensive research on a variety of applications, including Josephson junctions realized with bosonic gases confined in double-well potentials. Such systems have also been implemented experimentally,
allowing the observation of Josephson oscillations and macroscopic quantum self-trapping \cite{albiez2005direct,levy2007ac,doi:10.1126/science.1062612}. These systems are
commonly referred to as Bose–Josephson junctions (BJJs) \cite{pitaevskii2016bose}.
Owing to the high degree of tunability of their parameters, ultracold atomic systems have become an ideal testbed
for exploring many-body dynamics far from equilibrium, including the emergence of ergodicity and thermalization in
isolated quantum systems \cite{kinoshita2006quantum,hofferberth2007non,cheneau2012light,kaufman2016quantum,sinha2019chaos,PhysRevB.106.035123}. In recent years, particular attention has been devoted to the thermalization properties of BJJs and to the role played by chaotic dynamics when these systems are coupled to an external environment \cite{sinha2019chaos,sinha2019dissipative}. Such a coupling is often introduced because a single-species BJJ is an integrable system, and integrability
must be broken in order for chaos to emerge.
An alternative route to the breakdown of integrability is provided by interactions between multiple BJJs. Along
these lines, the case of two coupled BJJs composed of different bosonic species has already been investigated \cite{mondal2022classical}.
The three-species BJJ considered here represents a natural and significantly richer extension. The additional degree
of freedom enlarges the available phase space and gives rise to a more intricate dynamical landscape, potentially
supporting stronger chaotic behavior and novel thermalization mechanisms that remain largely unexplored.
This simple yet experimentally accessible model provides a bridge between the physics of Bose–Josephson junctions and the broader field of quantum thermalization in many-body systems. In this work, we investigate how the
presence of three distinct bosonic species affects ergodicity breaking and the emergence of nonthermal (scarred) eigenstates. Furthermore, the larger Hilbert-space dimension associated with the three-species system allows for a more
comprehensive analysis of quantum thermalization across separable, integrable, and chaotic regimes.
In particular, we find that the system exhibits thermal behavior in the integrable regime, while thermalization fails
in the fully separable limit. To demonstrate this, we assign an effective inverse temperature $\beta$ to each eigenstate
and examine whether the entanglement properties of excited states are consistent with the predictions of thermal
ensembles at the corresponding temperature.
The remainder of the paper is organized as follows. After introducing the model and the methods employed in this
work, we present our results in two parts. First, we analyze the onset of chaotic dynamics in multispecies BJJs
and investigate the associated quantum-scar phenomena that manifest as violations of ergodicity. Second, we focus on
specific parameter regimes in which the system becomes integrable or separable, allowing us to explore the possibility
of thermalization in the absence of chaos.
\section{Quantum Thermalization}
\subsection{Eigenstate Thermalization Hypotesis}
\label{ETH_text}
In quantum physics, one of the modern formulations of quantum thermalization is given by the Eigenstate Thermalization Hypothesis (ETH), according to which the reduced density matrix for a subsystem corresponding to any of the excited eigenstates is thermal \cite{garrison2018does}.
In other words, ETH states that every single excited energy eigenstate is typical in the thermodynamic sense, i.e., if, considering a local observable, the eigenstates themselves produce thermal expected values. Therefore, if the system satisfies ETH, \textit{every} excited eigenstate is already typical, not just a generic superposition of them, which accounts for almost all initial states, as ensured by normal typicality. Since each eigenstate is thermal, unitary evolution naturally brings any initial state to thermal equilibrium \cite{BALDOVIN20251}.\\
In particular, if ETH holds true, then, in the thermodynamic limit, the equal-time correlators of an appropriate class of operators, with respect to a finite energy density eigenstate $|n\rangle$ should be precisely equal to those derived from the canonical ensemble \cite{garrison2018does}, i.e.
\begin{equation}
\label{ETH-beta}
    \langle n |\hat O |n\rangle = \frac{tr(\hat Oe^{-\beta \hat H})}{tr(e^{-\beta \hat H})}
\end{equation}
where $\beta$ is chosen such that Eq.\eqref{ETH-beta} holds true when $\hat O = \hat H$, the Hamiltonian. We will use the notation $|n\rangle_\beta$ to denote an eigenstate whose energy density corresponds to temperature $\beta ^{-1}$.\\
In particular, these operators must belong to a subsystem A of the entire system such that $V_A<<V$ \footnote{There is some evidence that Eq.\eqref{ETH-beta} may also be valid if $V_A<\frac{V}{2}$ \cite{garrison2018does}.}.  If Eq. \eqref{ETH-beta} holds for \textit{all} operators within a subsystem A with these characteristics, we can equivalently state that \cite{nandkishore2015many}:
\begin{equation}
    \rho_A(|n\rangle_\beta)=\rho_{A,\text{can}}(\beta)
    \label{ETH_rho}
\end{equation}
where
\begin{equation}
    \begin{split}
        & \rho_A(|n\rangle_\beta) = tr_{\bar A} |n\rangle_{\beta\beta}\langle n|\\
        & \rho_{A,\text{can}}(\beta) = \frac{tr_{\bar A}(e^{-\beta \hat H})}{tr(e^{-\beta \hat H})}
    \end{split}
\end{equation}
with $\bar A$ being the complement of A.
\newline
As a consequence of this, the entanglement entropy on subsystem $A$, $S_A = Tr\{\hat\rho_A \ln \hat\rho_A\}$, built from a single high-energy eigenstate, $\hat\rho_A = Tr_{\bar A}\{|n\rangle_{\beta\beta}\langle n|\}$, not only scales extensively with the volume of A, but is essentially the thermal entropy of that subsystem at the corresponding temperature, $S_{th}(\beta)$ \cite{garrison2018does,PhysRevA.90.032103}. \\
While the subsystem formulation discussed above perfectly captures the thermal nature of individual eigenstates—effectively relying on the diagonal behavior of observables—the original microscopic formulation of ETH also addresses the dynamics of temporal relaxation towards such equilibrium. This is encapsulated in the standard full ansatz \cite{srednicki1994chaos,deutsch2018eigenstate}, which defines both the diagonal and off-diagonal matrix elements of a local observable $\hat{O}$ as:
\begin{equation}
\langle n| \hat{O}|m\rangle = \mathcal{O}(\bar{E}) \delta_{nm} + \Omega(\bar E)^{-1/2} f_O(\bar{E}, \omega) R_{nm}\end{equation}
where $\bar{E} = (E_n + E_m)/2$ and $\omega = E_n - E_m$. This indicates that the diagonal elements of the observable $\hat{O}$ in the energy basis, consistent with our previous discussion, resemble the microcanonical estimate—or the canonical one, assuming ensemble equivalence at temperature $\beta$. Meanwhile, the off-diagonal part is characterized by a smooth function $f_{O}(\bar{E}, \omega)$ that vanishes as the inverse square root of the density of states $\Omega$. The remaining factors $R_{nm}$ are tipically treated as independent, normally distributed random variables with zero mean and unit variance. \\
However, ETH as stated above has been challenged by a recent experiment, in which it was observed that some specific initial states in an array of strongly interacting ultracold Rydberg atoms do not thermalize and show long-lasting coherent oscillations, in contrast to the majority of initial states, which instead undergo ergodic temporal evolution in the same energy regime \cite{bernien2017probing}.\\
These atypical states in the ergodic regime lead to a weak breakdown of ergodicity, called \textit{Quantum Many-Body Scars} (QMBS) \cite{sinha2024classical}.\\
We speak of a weak breaking of ergodicity because, despite being strongly non-thermal, the scar states represent an infinitesimal fraction of Hilbert space and are immersed in a much larger sea of thermal eigenstates \cite{serbyn2021quantum}.
\subsection{Quantum Thermalization in integrable systems}
It is now clear that in classical mechanics, non-integrability and in particular chaos are neither necessary nor sufficient conditions for ergodicity and thermalization to occur on fast time scales \cite{BALDOVIN20251}.
This can be understood by citing Khinchin's ergodic theorem \cite{aleksandr1949mathematical}, according to which, in a high-dimensional system, the real key ingredient for achieving thermalization is the limit of a large number of bodies involved \textit{N}, regardless of the presence of chaos: thermal behaviour is only a matter of choosing the description of the system in terms of appropriate variables. For sufficiently large systems, it is sufficient to consider appropriate observables to ensure that the dynamic averages correspond to the averages of the corresponding ensemble and therefore have thermal behaviour. \\
However, historically, ETH originated as an extension of Berry's conjecture, according to which the high-energy eigenstates of a chaotic quantum system can be decomposed into random components (Fourier components or site amplitudes) with normal distribution \cite{berry1977regular,sinha2024classical}: the definition of ETH implicitly includes the condition of non-integrability of the system as a starting point for quantum thermalization.
Nonetheless, while ETH has been extensively tested in various interacting quantum systems, a rigorous proof establishing whether ETH is a necessary condition for thermalization is still lacking \cite{sinha2024classical}. Furthermore, whether or not non-integrability is required for thermalization is still a matter of debate in quantum mechanics. 
While recent years have seen an intensification in the study of thermalization in non-chaotic and integrable systems \cite{cattaneo2025thermalization, BALDOVIN20251}, a comprehensive framework that unifies these findings remains elusive. Significant progress has been made, for instance, investigations into global quantum quenches have identified a weak ETH scenario in separable systems \cite{biroli2010effect}.
On the other hand, large-scale analysis of eigenstate-to-eigenstate fluctuations of the reduced density matrix in integrable systems have revealed a structured thermal behavior, where fluctuations follow a Gaussian distribution with a standard deviation decaying as a power-law in the system's size \cite{alba2015eigenstate}, in clear contrast to the exponential decay typically observed in generic non-integrable systems. However, a critical gap remains in the systematic and unified characterization of these properties. \\
A vast body of literature explores ETH in integrable systems through the lens of the Generalized Gibbs Ensemble (GGE), which incorporates local integrals of motion into the density matrix to account for the constraints of integrability \cite{Essler_2016}. Numerical studies of quench dynamics in such systems consistently support GGE predictions \cite{PhysRevLett.98.050405,PhysRevLett.106.140405}. Typically, this form of integrability is analyzed via the Bethe Ansatz, which yields a set of local conserved charges scaling with the system size.
In this work, however, we focus on a different class of integrability. Our system is defined by a semiclassical Hamiltonian that,  for a specific set of parameters, is integrable in the sense of Liouville—implying, in our case, the existence of global, rather than local, conserved quantities—and characterized by Poissonian level spacing statistics. While it is formally possible to construct non-local integrals of motion for any quantum system using energy-state projectors, our focus lies on constants of motion that are globally defined functions of phase-space variables, possessing a clear semiclassical geometric interpretation.
Furthermore, it has been shown \cite{PhysRevB.94.245131} that even in non-interacting translational invariant quadratic systems—where an extensive number of local integrals of motion exist—the majority of eigenstates can still be described by a standard Gibbs ensemble characterized by a single Lagrange multiplier, $\beta$, a scenario often referred to as weak ETH. This work aims to extend this phenomenology beyond non-interacting quadratic models to an interacting integrable system. Additionally, we investigate how thermal-like behavior can emerge in systems that lack the characteristic fluctuations of quantum chaos.
To characterize this behavior, we use a framework where thermalization properties are intrinsically encoded in the assignment of a unique inverse temperature $\beta$ to each individual eigenstate.
We evaluate the validity of this thermal mapping through a qualitative analysis, examining the extent to which the entanglement properties of excited states replicate the theoretical predictions derived from their assigned temperatures. This is achieved by employing an entanglement entropy analysis following an approach that is distinct from prior studies on interacting integrable systems \cite{alba2015eigenstate}.\\
Furthermore, to provide a comprehensive and unified framework, we achieve simultaneously a broad comparison across the separable, integrable, and chaotic regimes within the same physical framework.
Crucially, the interacting integrable regime of the system used in this study was previously undocumented; its derivation and proof of integrability represent another original contribution of this work, enabling a controlled study across all three dynamical regimes.
\section{Models and Methods}
When two Bose-Einstein condensates are trapped in a double-well potential and the barrier is large enough to ensure a weak connection between both condensates on each side of the trap, the quantum phase difference will cause rapid oscillatory tunnelling of atoms through the potential barrier, called the Josephson effect \cite{leggett2001bose}. These kind of systems are called Bose-Josephson Junction (BJJ).\\
The BJJ of a mixture of three species of ultra-cold bosons with equal population $N$ of each species can be described using the Bose-Hubbard formalism \cite{mele2012spin}:
\begin{equation}
         \hat H = -\frac{J}{2}\sum_{\alpha = 1,2,3}(\hat a_{\alpha L}^\dagger\hat a_{\alpha R} +\hat a_{\alpha R}^\dagger\hat a_{\alpha L}) 
         + \frac{U}{2N}\left[\hat N_L(\hat N_L-1)+ \hat N_R(\hat N_R-1)\right]
         +\sum_{j = L,R}\epsilon_j\hat N_j
\end{equation}
where 
\begin{equation}
    \hat N_j = \sum_{\alpha = 1,2,3}\hat a_{\alpha j}^\dagger\hat a_{\alpha j}
\end{equation}
Here, $J$ is the tunneling coupling between sites, $U$ is the interaction term\footnote{The Bose-Josephson Junction is typically described by the Hamiltonian \cite{PhysRevLett.81.3108}:
\begin{equation}
     \hat H = -J\sum_{\alpha = 1,2,3}(\hat a_{\alpha L}^\dagger\hat a_{\alpha R} +\hat a_{\alpha R}^\dagger\hat a_{\alpha L}) 
     + \frac{U}{2}\left[\hat N_L(\hat N_L-1)+ \hat N_R(\hat N_R-1)\right]
     +\sum_{j = L,R}\epsilon_j\hat N_j
\end{equation}
Here, we adopt the standard mean-field scaling for the interaction ($U \to U/N$) to maintain a well-defined thermodynamic limit. This scaling allows us to systematically compare the macroscopic classical dynamics ($N \to \infty$) with the exact quantum evolution for finite $N$.}.
To build a more realistic model, different couplings for intra-particle and inter-particle interactions can be considered. Neglecting all constant terms that cause only a rigid displacement of energy, the Hamiltonian becomes:
\begin{equation}
    \label{spinorial_BJJ}
         \hat H = -\sum_{\alpha = 1,2,3}[\hat a_{\alpha L}^\dagger\hat a_{\alpha R} +\hat a_{\alpha R}^\dagger\hat a_{\alpha L}
         + \frac{U}{2N}(\hat n_{\alpha L}(\hat n_{\alpha L}-1)+ \hat n_{\alpha R}(\hat n_{\alpha R}-1))
         + \frac{V}{2N}\sum_{\alpha'\ne\alpha}(\hat n_{\alpha L}\hat n_{\alpha' L} + \hat n_{\alpha R}\hat n_{\alpha' R})]
\end{equation}
We set $\hbar = 1$ and scale energy  by \textit{J} and time by \textit{1/J}.
\label{Models and Methods}
The Hamiltonian in Eq.\eqref{spinorial_BJJ}, can be written as an effective Hamiltonian describing three interacting large spins:
\begin{equation}
    \label{H_BJJ_Schwin}
    \hat H = \sum_{\alpha = 1,2,3}\left(-\hat S_{\alpha x}+ \frac{U}{2S}\hat S_{\alpha z}^2 + \frac{V}{S}\sum_{\alpha'<\alpha}\hat S_{\alpha z} \hat S_{\alpha' z}\right)
\end{equation}
where for each species with $S=N/2$ 
\begin{equation}
    \begin{split}
        \hat S_{\alpha x} &= (\hat a_{\alpha L}^\dagger\hat a_{\alpha R} +\hat a_{\alpha R}^\dagger\hat a_{\alpha L})/2\\
        \hat S_{\alpha z} &= (\hat n_{\alpha L} - \hat n_{\alpha R})/2
    \end{split}
\end{equation}
using the Schwinger-Boson representation \cite{milburn1997quantum}.
We can see that, in the absence of the term $V$, the Hamiltonian is reduced to being separable, constructed as the sum of integrable Hamiltonians
\begin{equation}
\label{sep_hamil}
    H = H_{BJJ,\alpha = 1}+H_{BJJ,\alpha = 2}+H_{BJJ,\alpha = 3}
\end{equation}
with
\begin{equation}
	H_{BJJ,\alpha} = -\hat S_{\alpha x}+ \frac{U}{2S}\hat S_{\alpha z}^2 .
\end{equation}
The term $V$ in Eq.\eqref{H_BJJ_Schwin} makes the system non-separable \footnote{In the special case in which $V = 2U $ we recover the original structure of the Hamiltonian in eq.\eqref{spinorial_BJJ}. }, which can also compromise the integrability of the total system in the semiclassical regime. 
\subsection{Semiclassical limit}
The semiclassical limit is recovered through the limit $\textstyle{N>>1}$, where we can use the approximation on Bose-Hubbard Hamiltonian in which
\begin{equation}
    \hat a_j\sim\sqrt{N_j}e^{i\phi_j}
\end{equation}
where $N_j$ and $\phi_j$ become conjugate variables.
Considering the canonical change of variables
\begin{equation}
    \begin{split}
        z &= N_L-N_R \\
        \delta \phi &= \frac{\phi_L-\phi_R}{2}\\
        N &= N_L+N_R \\
        \Phi &= \frac{\phi_L+\phi_R}{2}\\
    \end{split}
\end{equation}
the Hamiltonian of the system is reduced to:
\begin{equation}
    \label{semicalssical_BJJ}
        \frac{H}{N/2} =\tilde{H}=-\sum_{\alpha = 1,2,3}(\sqrt{1-\tilde{z}_\alpha^2}\cos( \tilde\phi_\alpha)
        +\frac{{U}}{2}\tilde{z}_\alpha^2 + V\sum_{\alpha'<\alpha}\tilde z_\alpha\tilde z_{\alpha'})
\end{equation}
with
\begin{equation}
\begin{cases}
    \tilde z_\alpha = z_\alpha/N\\
    \tilde\phi_\alpha = 2\delta\phi_\alpha
\end{cases}
\end{equation}
This limit coincides with that done considering large magnitude of spin $S>>1$, for which the effective spin operators can be treated as components of the classical spin vector \\
$\vec{S_\alpha}=S(\sin\theta_\alpha\cos\phi_\alpha,\sin\theta_\alpha\sin\phi_\alpha,\cos \theta_\alpha)$ where 
\begin{equation}
\begin{split}
    \tilde\phi_\alpha &= \phi_\alpha\\
    \tilde z_\alpha &= \cos\theta_\alpha  .
\end{split}
\end{equation}
\section{Results}
\subsection{Quantum Chaos}
We wrote the Hamiltonian in Eq.\eqref{H_BJJ_Schwin} in basis $|i\rangle =|m_{z1},m_{z2},m_{z3}\rangle$, with $m_{zj}$ the eigenvalues of $\hat S_{zj}$, and then diagonalized it to compute the eigenvalues $E_n$ and eigenfunctions $|n\rangle = \sum_{i=1}^D c_n^i|i\rangle$. Therefore, the dimension of the Hilbert space is $ D = (2S+1)^3$.\\
To investigate the signature of chaos at the quantum level, we studied the spectral properties of the system \cite{haake1991quantum}. We sorted the eigenvalues $E_n$ belonging to a particular symmetry sector of the Hamiltonian, and compute the unfolded level spacing distribution $P(\delta E)$, where $\delta E_n = \frac{E_{n+1}-E_n}{D_n}$ and $D_n$ is the average level spacing in a small vicinity of $E_n$
(See Appendix \ref{appendix_symmetry} and \ref{appendix_unfolding}).  If the semiclassical limit of the system is chaotic, according to Bohigas-Giannoni-Schmit (BGS) conjecture \cite{bohigas1984characterization}, the level spacing distribution should follow Wigner-Dyson statistics, $P(\delta E) =(\pi\delta E/2)\exp(-\pi\delta E^2/4)$, corresponding to Gaussian orthogonal ensemble \footnote{The choice of Gaussian random matrix ensemble is determined by the underlying symmetries of the system. If the Hamiltonian is Hermitian with complex entries and no further symmetry constraints, the appropriate ensemble is the Gaussian Unitary Ensemble (GUE). In contrast, if the Hamiltonian is real and symmetric, the system corresponds to the Gaussian Orthogonal Ensemble (GOE). } (GOE) statistics belonging to the Random Matrix Theory (RMT); whereas Poissonian statistics,  $P(\delta E) = \exp(-\delta E)$ can be observed in system with a integrable semiclassical analogous \cite{berry1977level,haake1991quantum}. 
\begin{figure}[ht]
	\centering
	{\includegraphics[width=0.4\textwidth]{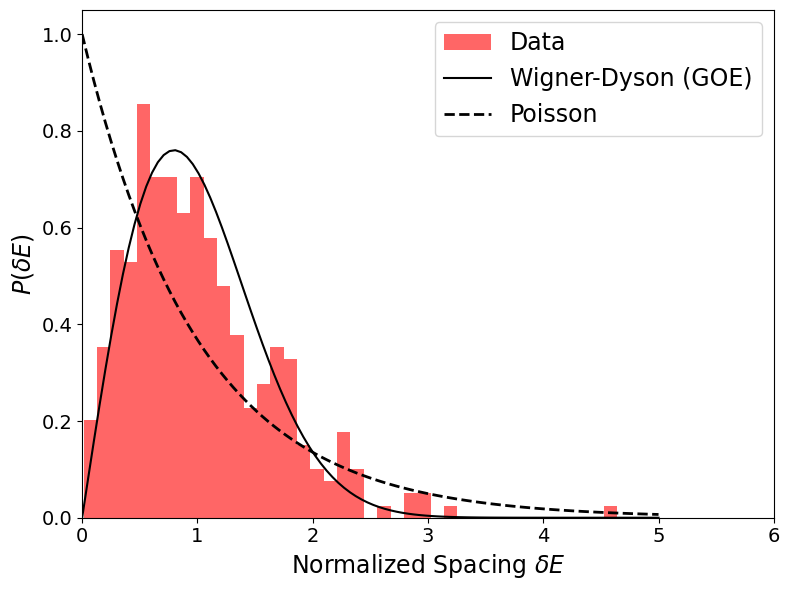}}
	\caption{Unfolded level spacing distribution with parameters $U = 1$, $V=2$ and $S_\alpha = 8$. The dashed and the continuous lines give respectively the Poisson and Wigner-Dyson distribution.}
	\label{WD_U_1V_2}
\end{figure}
As we mentioned in Sect. \ref{Models and Methods}, the $V$-term in Eq.\eqref{H_BJJ_Schwin} can induce a non-integrable perturbation in the Hamiltonian, and it may be interesting to study whether and where the system exhibits chaotic features for certain combinations of parameters $(U,V)$. We show in Fig.\ref{WD_U_1V_2}that it is indeed possible to achieve the Wigner-Dyson distribution for certain values of $(U,V)$ and therefore quantum chaos for this system.\\
Thus, we computed the average level spacing ratio \cite{oganesyan2007localization},
\begin{equation}
	\langle r\rangle = \langle \min (\delta E_n,\delta E_{n+1})/\max (\delta E_n,\delta E_{n+1})\rangle,
\end{equation} 
as an indicator of quantum chaoticity, because the number it returns is tabulated: for the Poisson distribution, $\langle r\rangle\sim0.386$, and for GOE, $\langle r \rangle\sim0.530$ \cite{atas2013distribution}.

\begin{figure}[ht]
	\centering
	{\includegraphics[width=0.5\textwidth]{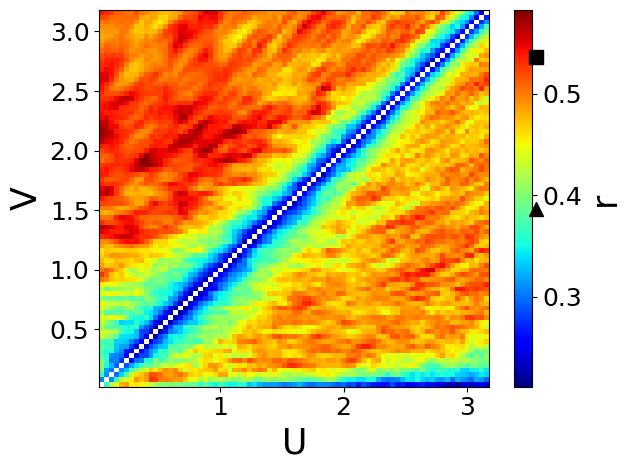}}
	\caption{Averaged level spacing ratio in the $(U,V)$-parameter space with $S_\alpha = 5$. The black triangle and square in the colorbar indicate the theoretical values of $\langle r \rangle$ for the Poisson and Wigner-Dyson distributions, respectively. }
	\label{UVplane}
\end{figure}

Fig.\ref{UVplane} shows that the system presents a wide region in $(U,V)-$parameter space in which the \textit{degree of chaos} is very high according the average level spacing indicator \footnote{The colorbar in Fig.\ref{UVplane} is not bounded from 0.386 to 0.530 because of the finite size effect. For a wider discussion see the Appendix \ref{appendix_finite}.}. This phenomenology is also present in the lower-dimensional model that considers only two different species \cite{mondal2022classical}.
\subsection{Quantum Scars}
\begin{figure}[ht]
 	\centering
 	{\includegraphics[width=0.65\textwidth]{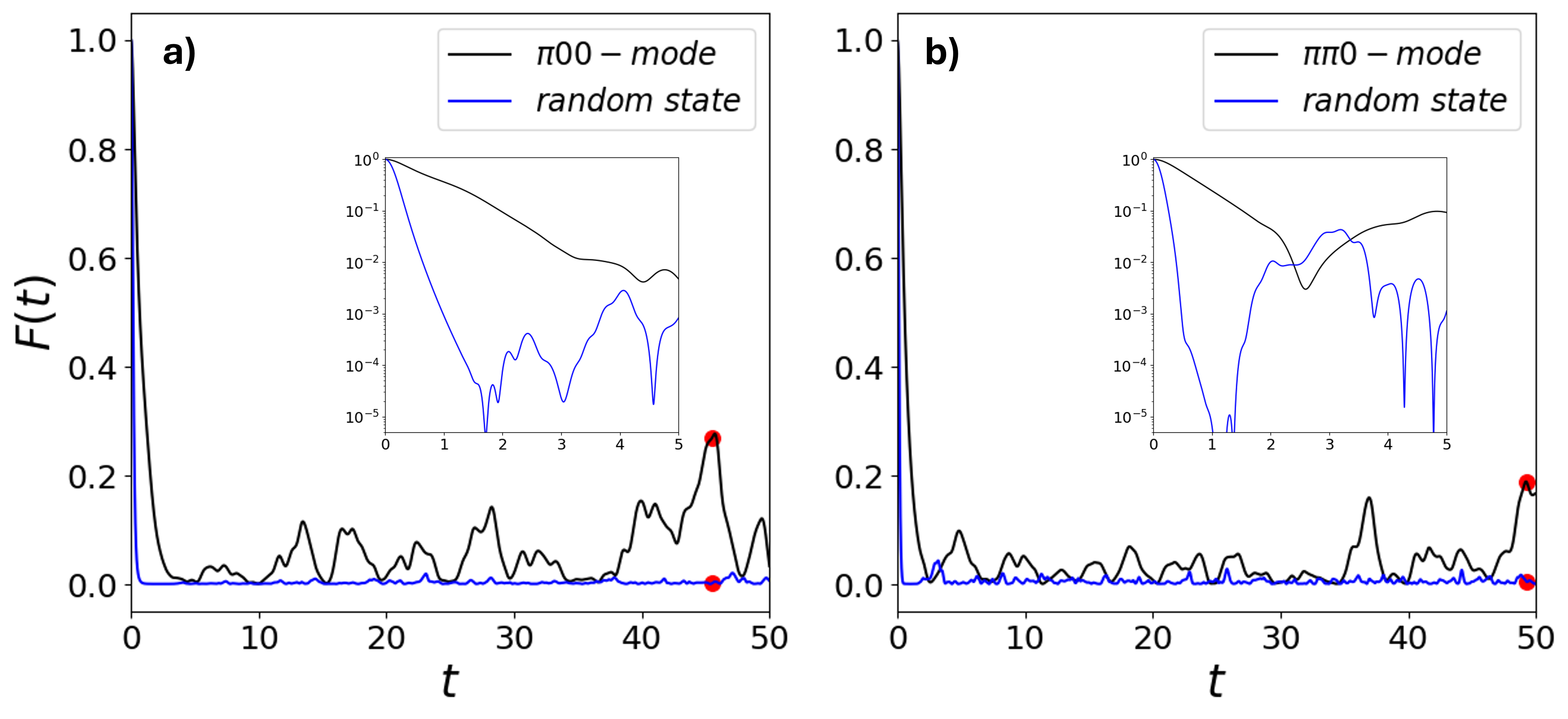}}
 	\caption{Survival probability for both the scar states $-$ in panel a) the '$\pi 0 0$-mode' and in b) the '$\pi\pi 0$-mode' $-$ and the corresponding random states. Red markers indicate the times of maximum self-overlap for the scar state relative to $t=0$ (see appendix \ref{appendix_husimi} for details). The simulation was carried out with every $S_i$ = 6 and the parameters $U = 1, V = 2$. } 
 	\label{Survival}
 \end{figure}

\paragraph{Semiclassical model} 
In the study of quantum thermalization, collective oscillations are studied so much because their relevance in the breaking ergodicity phenomena related to quantum scarring. In systems such as the one under study, we can recognise these collective modes as fixed points present in the dynamic system constructed from the classical equations of motion:
\begin{equation}
	\label{dyn_sys_cl}
	\begin{split}
		\dot{\tilde z}_i &= -\sqrt{1-\tilde z_i^2}\sin\tilde\phi_i\\
		\dot{\tilde{\phi_i}} &= \frac{\tilde z_i\cos\tilde\phi_i}{\sqrt{1-\tilde z_i^2}} + U\tilde z_i + V (\tilde z_{i+1}+\tilde z_{i-1})
	\end{split}
\end{equation}
where $i = 0,1,2$, and periodically  $i = 3 = 0$ and $i = 0 = 2$.\\
It can be easly seen that setting every \footnote{For ease of reading, we omit the formalism  $\tilde z$ and $\tilde \phi$ and use simply $z$ and $\phi$.} $z_i = 0$, the fixed points are restricted to $\phi_i = 0,\pi$.
\
Compared with the two species case \cite{mondal2022classical}, the complexity of our phase diagrams allows for a higher number of fixed points.
It is evident from calculating the energy of the resulting fixed points that the only way to have a semiclassical energy that is distant from the boundaries of the quantum energy spectrum — the scenario in which ETH should be applicable — is to have one or two $\phi_i = \pi$. We will call these collective oscillations '$\pi\pi0$-mode' and '$\pi00$-mode' because they are collective modes in which the difference of $\phi_i$ can be 0 or $\pi$, depending on the chosen fixed point. These two fixed points might be suitable candidates for recovering scarred states because it was shown that the phenomenon of quantum scarring can be closely related to unstable collective oscillations \cite{serbyn2021quantum}. We set $V=2U=2$ to ensure the instability of all classical fixed points (see Appendix \ref{appendix_fixed} for explicit computation).\\
\paragraph{Quantum model}
The quantum states corresponding to these classical points can be described with a representation as coherent states, since in them the packet has minimal uncertainty around the point itself. Thanks to the Schwinger-boson representation, the Hamiltonian is reduced to a system of interacting spins, for which there is a coherent analytical description of the spin states \cite{radcliffe1971some}:
\begin{equation}
		|z,\phi\rangle=\left(\frac{1+z}{2}\right)^S\exp\left(\sqrt{\frac{1-z}{1+z}}e^{i\phi}\hat S_{-}\right)|S,S\rangle.
\end{equation}
In such a way, the  quantum analogous of '$\pi\pi0$-mode' and '$\pi00$-mode' are naturally the states 
\begin{equation}
\label{coherent_states_pi}
	\begin{split}
		|\psi(0)\rangle_{\pi 00} &=\frac{1}{\sqrt 3}\left(|0\rangle|0\rangle|\pi \rangle + |0\rangle|\pi\rangle|0 \rangle + |\pi\rangle|0\rangle|0 \rangle\right)\\
		|\psi(0)\rangle_{\pi \pi0} &=\frac{1}{\sqrt 3 }\left(|0\rangle|\pi\rangle|\pi \rangle + |\pi\rangle|0\rangle|\pi \rangle + |\pi\rangle|\pi\rangle|0 \rangle\right)
	\end{split}
\end{equation}
where we have dropped the $z$-variable in $|z,\phi\rangle$, writing simply $|\phi\rangle$, since $z=0$ for all modes considered here.\\
In order to analyze the degree of scarring of these candidate states, we examine the \textit{survival probability}
\begin{equation}
	F(t)=|\langle\psi(0)|\psi(t)\rangle|^2
\end{equation}
which measures the extent to which the time-evolved state retains memory of the initial state. A slower decay or persistent revivals in $F(t)$ indicate a substantial overlap with $|\psi(0)\rangle$ over time. This behavior is often associated with non-ergodic dynamics and therefore, in this case, with the presence of quantum many-body scar states, in contrast to the rapid decay expected in typical thermalizing systems.\\
We initialize the system in a pure coherent state corresponding either to the predicted scar states (as described in Eq.\eqref{coherent_states_pi}) or to a random set of parameters $\{z_i,\phi_i\}$ sampled at the same semiclassical energy as the corresponding scar.
Fig. \ref{Survival} displays the survival probability for these states. It is evident that the scar states, unlike the arbitrary state in which a decline in function is observed, do not lose information on the initial data, maintaining a large time-dependent overlap. Furthermore, the semi-logarithmic inset reveals that the characteristic exponential decay time for scar states is at least one order of magnitude longer than that of a generic coherent state as expected \cite{serbyn2021quantum}. We refer the reader to Appendix \ref{appendix_husimi} for a visualization of the quantum phase space, further illustrating the behavior discussed here.\\

\subsection{Thermalization and integrability}
By analyzing the Hamiltonian in Eq. \eqref{H_BJJ_Schwin}, we identify specific parameter regimes where the system is integrable, as illustrated in fig. \ref{UVplane}. We focus our analysis on two distinct cases: the trivial separable case ($V=0$) and the non-trivial integrable case ($U=V$). In the first case ($V=0$), the Hamiltonian is separable, and each $H_{BJJ,\alpha}$ serves as a constant of motion. Due to the separability of the Hamiltonian, the corresponding eigenstates will be disentangled in the single-spin basis, and for this reason we do not expect the thermalization property for the eigenstates themselves. As shown in the appendix \ref{appendix_separability}, introducing an integrability-breaking term by increasing $V$ triggers a transition toward thermal behavior, consistent with established results in the literature \cite{deutsch2018eigenstate}. Indeed, the breaking of integrability is typically associated with a transition toward an ETH-compliant regime.
In contrast, we isolate the second case, where $U=V$. Under these conditions, the Hamiltonian reduces to:
\begin{equation}
	H = - \sum_{\alpha = 1,2,3} S_{\alpha x}+ \frac{U}{2S} \left(\sum_{\alpha = 1,2,3}  S_{\alpha z}\right)^2
\end{equation}
We rigorously establish in Appendix \ref{appendix_integrable} that this system remains integrable due to the emergence of additional constants of motion —specifically $(\vec S_1+\vec S_2+\vec S_3)^2$ and  $(\vec S_1+\vec S_2)^2$. Crucially, however, the interspecies interaction in this regime induces significant entanglement in the single-spin basis, despite the system’s overall integrability. This observation forms the core of this work: we investigate whether such interaction-induced entanglement is sufficient to confer thermalization properties upon the majority of the individual eigenstates, or if the breaking of integrability remains a necessary prerequisite.

\begin{figure}[ht]
	\centering	\includegraphics[width=1.1\textwidth]{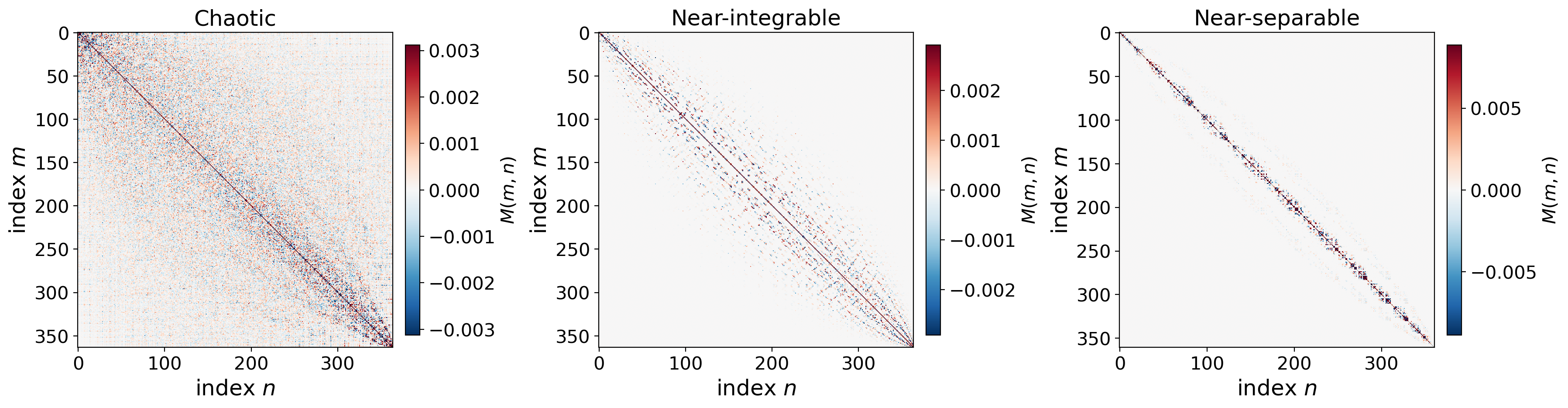}
	\hfill
	\caption{The off-diagonal elements $M_{nm}$ of the observable $\hat \rho ^{(1)}(\beta)$. From left to right: the non-integrable regime ($V=2U=2$), the integrable interacting regime ($U\sim V=1$), and the separable regime ($V\sim0$). Here $N = 2 S = 12 $.}
	\label{ETH_M}
\end{figure}

\begin{figure}[ht]
	\centering	\includegraphics[width=0.6\textwidth]{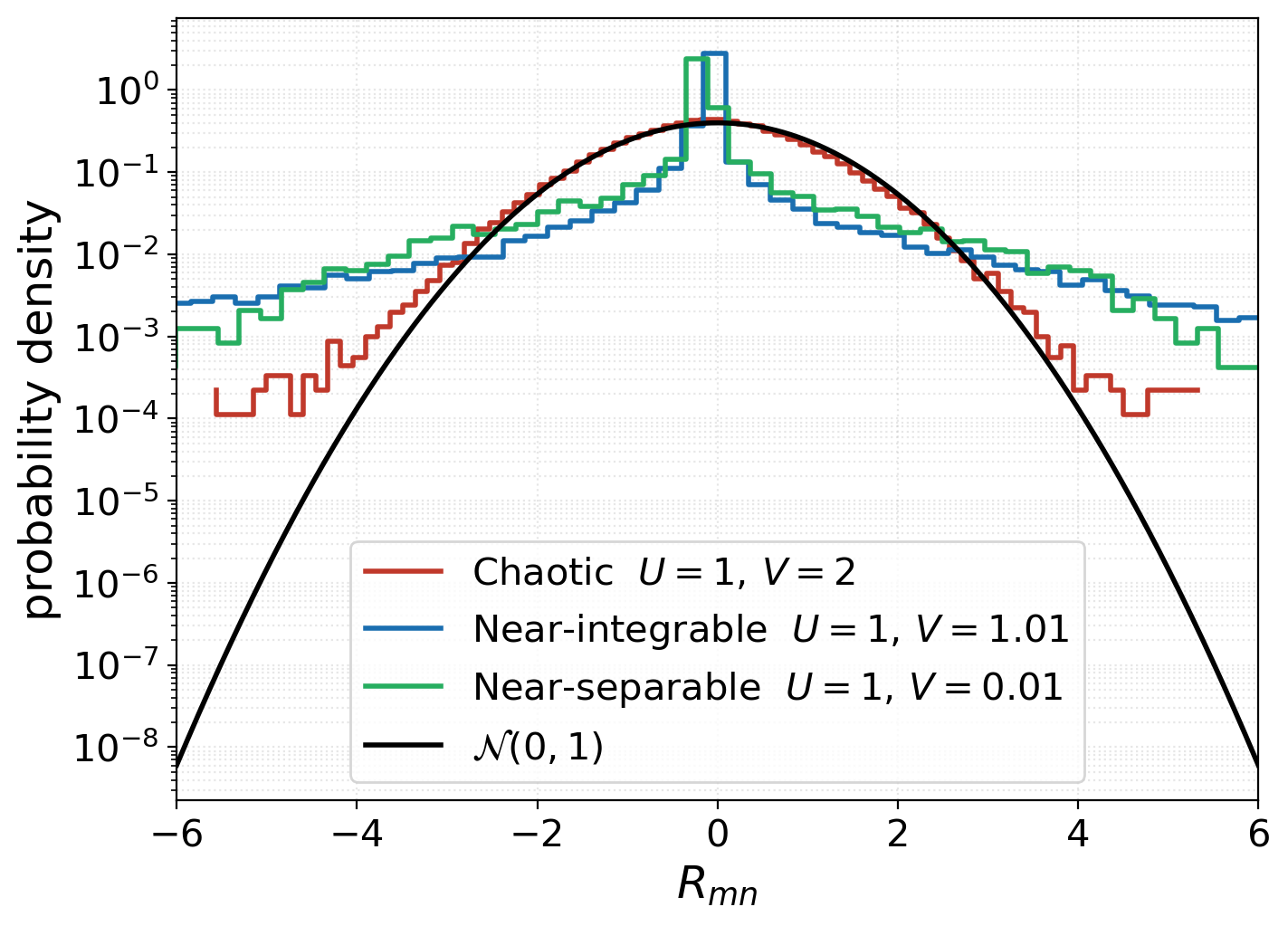}
	\hfill
	\caption{The ETH $R_{nm}$ distribution for all the regimes. Here $N = 2 S = 12 $.}
	\label{ETH_R}
\end{figure}

Trying to answer this question, we analyzed the Entanglement Entropy (EE) of each individual eigenstate considering a reduced density matrix as defined in Eq.\eqref{rhored} on a single species, in the spirit of what is stated in Section \ref{ETH_text}.
According to this, if ETH is valid, the EE trend of individual eigenstates, 
should follow the thermal prediction, allowing us to define an effective temperature $\beta^{-1}$ for each eigenstate $|n\rangle_\beta$.\\
\begin{figure}[h]
	\centering	\includegraphics[width=0.6\textwidth]{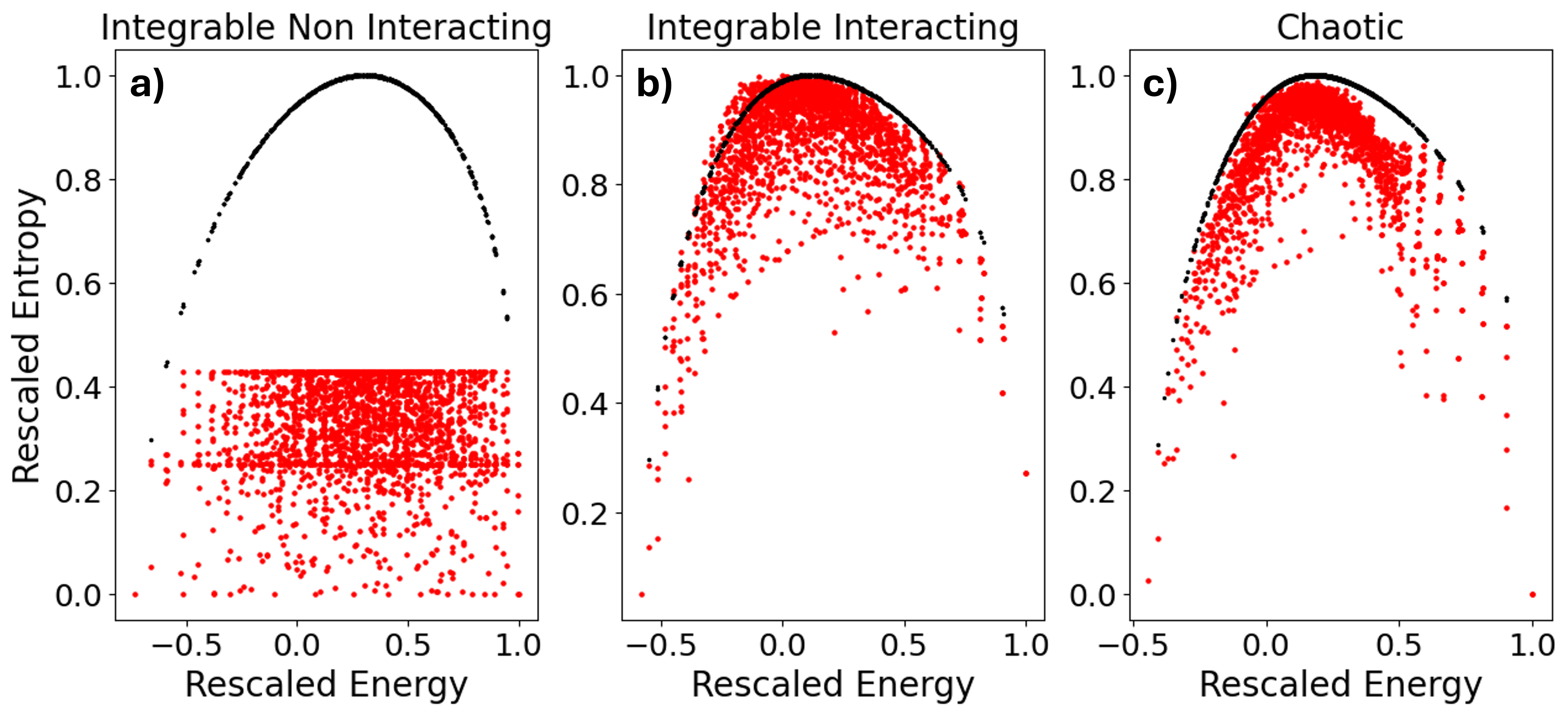}
	\hfill
	\caption{Rescaled EE computed for every eigenstate in function of the corresponding rescaled eigenenergy using  $S_i$ = 6. EE refers to a single spin and the theoretical thermal curve is obtained using equation \eqref{ETH_rho}. Panel a) represents the separable case with $V = 0$, b) the integrable but interacting one with $U = V$, and c) the chaotic regime with $U = 1$ and $V=2$. The rescaled entropy indicates the entropy divided by the maximum entropy that can be obtained: $\ln (2S+1)$. The rescaled energy indicates the energy divided by the maximum eigenvalues of the system.}
	\label{EE}
\end{figure}
Figure \ref{EE} compares the rescaled EE —normalized by the maximum possible value $\ln(2S+1)$— with the theoretical thermal prediction from Eq.\eqref{ETH_rho}. This comparison is shown for three distinct regimes: the non-integrable system, the integrable interacting case ($U = V$), and the non-interacting limit ($V = 0$). We can observe that, regardless of the integrability of the system, in interacting systems the EE follows the theoretical trend predicted by ETH \footnote{The large fluctuations in the entanglement entropy (EE) shown in Fig.~\ref{EE} stem from finite-size effects. Because the BJJ Hamiltonian reduces to an effective lower-dimensional system, it is not a true many-body system. Consequently, the Hilbert space dimension grows polynomially rather than exponentially with particle number, following $D = (2S+1)^3 = (N + 1)^3 \sim N^3$. The system-size dependence of the EE is detailed in Appendix~\ref{appendix_EE_scaling} for both the integrable and non-integrable regimes.}. 
On the contrary, in the case of no interaction, the ETH prediction is not reproduced by the EE of the individual pure eigenstates\footnote{In Fig. \ref{EE}, the entanglement entropy is computed for all eigenstates without resolving symmetry sectors. For a detailed analysis of how symmetry sector decomposition influences the degeneracy and the resulting EE, see Appendix \ref{appendix_EE_sectors}.}. \\
As stated in \ref{ETH_text}, the standard formulation of ETH is based on matrix elements of local observables. By this ansatz, the remaining factors in the off-diagonal part of matrix element in energy basis $R_{nm}$ are treated as independent, normally distributed random variables with zero mean and unit variance - a conceptual constraint inherited from Random Matrix Theory. Despite it is generally valid for chaotic systems, where the connection to RMT is explicit, it is typically not true for integrable systems, which exhibit Poisson level statistics \cite{PhysRevE.100.062134}. Here, we investigate if this feature are really necessary to require thermalization properties.\\
For this analysis, we choose as our observable the single-site reduced density matrix of an excited eigenstate
\footnote{To ensure the robustness of our conclusions, we repeated the analysis across various excited eigenstates corresponding to different values of EE obtaining the same qualitative behavior. } 
located at the center of the spectrum ($\beta \sim 0$):\begin{equation}\hat{\rho}^{(1)}(\beta) = \text{Tr}_{2,3} { |n\rangle_{\beta\beta} \langle n| }\end{equation}
Figure \ref{ETH_M} presents a colormap of the off-diagonal elements of the reduced density matrix, which highlights a distinct qualitative transition across the three regimes. To quantify these differences, Fig. \ref{ETH_R} shows the distribution of the normalized off-diagonal matrix elements $R_{nm}$, obtained by diagonalizing the Hamiltonian within specific symmetry sectors. As expected, the non-integrable case ($V=2U$) exhibits the Gaussian fluctuations characteristic of Random Matrix Theory. Conversely, both the integrable interacting regime ($U=V$) and the separable regime ($V=0$) deviate from this behavior, confirming that the system lacks the RMT-like fluctuations typically associated with quantum chaos. Yet, the integrable interacting case remains thermal when evaluated through its entanglement entropy —a result reflecting thermalization at the level of individual eigenstates. This behavior implies that the diagonal elements of any local operator acting on a single-spin subsystem converge to the expected microcanonical (or canonical) thermal values. Thus, while the system fails to satisfy the off-diagonal RMT criteria, the diagonal part of the ETH ansatz is robustly satisfied. Consequently, RMT-like fluctuations are not a necessary prerequisite for thermalization; 
this provides further evidence that, in our integrable model, interaction-induced entanglement is sufficient to recover thermodynamic behavior.
\section{Summary and Discussion}
In this work, we investigated the chaotic dynamics and thermalization properties of an isolated quantum system: a three-species Bose--Josephson junction. Although the system exhibits quantum-chaotic behavior over a large portion of its parameter space, as quantified by the mean level-spacing ratio, we identified distinct deviations from ergodicity. Through a linear stability analysis of the corresponding semiclassical dynamics, we located unstable fixed points associated with athermal behavior and confirmed their quantum counterpart through the presence of scarred eigenstates.\\
The system also displays an integrable limit in the semiclassical limit, providing a unique framework that bridges previous studies of separable systems \cite{biroli2010effect} and interacting integrable systems \cite{alba2015eigenstate}. By assigning an effective inverse temperature $\beta$ to individual eigenstates 
we obtained a detailed state-by-state characterization of thermal properties. Our results show that the entanglement entropy of excited eigenstates closely follows the predictions of a thermal ensemble in both the interacting integrable and chaotic regimes. In contrast, this correspondence breaks down in the separable limit and remains strongly suppressed in its vicinity. 
These findings extend the weak ETH phenomenology out of the properties of quadratic systems to an integrable interacting case and suggest that Random Matrix behavior is not a prerequisite for thermalization of typical eigenstates; rather, interacting integrable systems can exhibit genuine thermal behavior in an experimentally accessible manner.
Notably, the proposed system can be realized with current ultracold-atom platforms \cite{taglieber2008quantum}, making our predictions directly amenable to experimental verification.

\section*{Acknowledgements}
Authors are grateful to Angelo Russomanno, Andrea Trombettoni and Lorenzo Piroli for fruitful discussions and comments on the work.


\paragraph{Funding information}
S.C. acknowledges funding from by the European Union
- NextGenerationEU through the Italian Ministry of University and Research (MUR), PNRR Mission 4 Component 2 Investment 1.3, project PE00000023 (NQSTI), Spoke 9, under the Cascade Call CUP E63C22002180006. F.D.M. and S.P. acknowledge support from Istituto Nazionale di Fisica Nucleare (INFN) through the QUANTUM project. 

\begin{appendix}

\section{Symmetry sectors}
\label{appendix_symmetry}
The eigenvectors and corresponding eigenvalues can be classified into the appropriate symmetry sectors defined by the symmetries of the Hamiltonian, written as operators that leave the Hamiltonian invariant under their action. This classification is essential because eigenvalues belonging to different symmetry sectors are statistically uncorrelated, and mixing levels from different sectors could therefore artificially influence the distribution of the interval between levels towards Poisson statistics \cite{d2016quantum}. \\
In the Hamiltonian in Eq.\eqref{H_BJJ_Schwin}, the eigenvalues can be divided first into even and odd sector of parity operator, which can be represented as
\begin{equation}
	\hat \Pi=e^{i\pi(\hat S_{x,1}+\hat S_{x,2}+\hat S_{x,3})}
\end{equation}
and then discriminated by spin exchange operators
\begin{equation}
	\hat O_1 , \hat O_{TOT} = \hat O_1 + \hat O_2 + \hat O_3
\end{equation}
such that
\begin{equation}
	\begin{split}
		\langle m_{3},m_{1},m_{2}|\hat O_1|m_{1},m_{2},m_{3}\rangle=1\\
		\langle m_{2},m_{3},m_{1}|\hat O_2|m_{2},m_{1},m_{3}\rangle =1\\
		\langle m_{1},m_{2},m_{3}|\hat O_3|m_{1},m_{2},m_{3}\rangle =1\\
	\end{split}
\end{equation}
and 
\begin{equation}
	[\hat O_1, \hat O_{TOT}] = 0
\end{equation}
where $m_{\sigma}$ are the quantum numbers of $\hat S_{z,\sigma}$.

\section{Unfolding procedure}
\label{appendix_unfolding}
The relationship between random matrices and complex quantum systems, as initially intuited by Wigner, relies on restricting attention to a narrow energy window in which the density of states is approximately constant. Within such a window, the Hamiltonian, when represented in a generic basis, can effectively be modeled as a random matrix. Therefore, in order to correctly use the BGS conjecture, care must be taken in how the spectrum of the system is analyzed. It is not sufficient to calculate the bare spectrum by simply diagonalizing the Hamiltonian. In fact, the universal spectral behavior given by the Wigner-Dyson distribution in chaotic quantum systems is only local in the spectrum: it breaks down for correlations involving many levels.\\
One way to see this universality is to magnify the spectrum so that the mean spacing of levels becomes the unity. Because the level spacing in a system with $N$ freedoms is of the order of $h^N$, we need of a magnification factor of the order of $h^{-N}$ \cite{berry1987bakerian}.\\
Operationally, the procedure for scaling energy levels is as follows:\\
let us introduce the cumulative density of states \cite{berry1977regular}  
\begin{equation}
	\begin{split}
		N(E) &=\int_{-\infty}^{E}P(E')dE' =\sum_{i}\Theta(E-E_i)\\
		&=N_{smooth}(E)+N_{fluct}(E)
	\end{split}
\end{equation}
where $P(E)=dN/dE$ is the local density of state, $N_{smooth}(E)$ incorporates the trend of the function with a certain degree of smoothness, and $N_{fluct}(E)$ describes the fluctuations around the uniform part. Then we propose the subsequent substitution \cite{santos2010onset}:
\begin{equation}
	E_j\rightarrow x_j=N_{smooth}(E_j)\hspace{0.05in}
\end{equation}
From conservation of probability we have:
\begin{equation}
	\label{unfolding}
	\begin{split}
		&\rho(x)dx = P(E)dE \\\implies
		\rho(x)&=\frac{P(E)}{dx/dE}\sim\frac{P(E)}{P(E)} = 1 
	\end{split}
\end{equation}
where we used that $dN_{smooth}(E)/dE\sim dN(E)/dE=P(E)$. For this we computed the unfolded level spacing distribution using $\delta E_n = \frac{E_{n+1}-E_n}{D_n}$ and $D_n = 1/(dN_{smooth}(E_n)/dE_n)$ is the average level spacing in a small vicinity of $E_n$.\\
We avoid the naive substitution
\[
\quad E_j \;\to\; x_j = N(E_j),
\]
because although this formally enforces a mean level spacing of 1, since for every \(i\),
\begin{equation}
	\Theta(E_{i+1}) - \Theta(E_i) = 1\hspace{0.05in}
\end{equation}
it trivializes the spacing distribution (all spacings become identical), rendering any statistical analysis meaningless.
Generally, the smooth part of the cumulative function is computed numerically with a polynomial fit of the total cumulative function. In this work, we used a polynomial fit for the cumulative function of \textit{7th} order, and discarding the eigenvalues at the edges of the spectrum by rejecting the first and last 10\%
from the total to consider only the bulk of the spectrum.
\section{Finite Size Effects}
\label{appendix_finite}
\begin{figure}[ht]
	\centering
	{\includegraphics[width=0.5\textwidth]{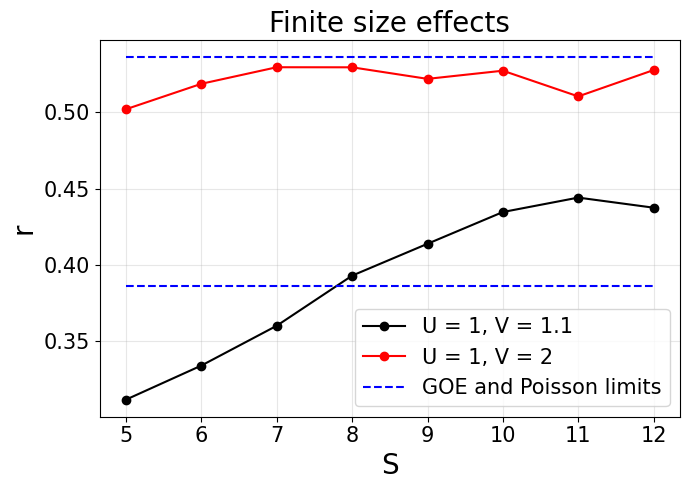}}
	\caption{Average level spacing ratio as a function of the magnitude of each spin, and so the number of particles $N = 2S$. The plots are in near-integrable case $U = 1$, $V=1.1$ (black), and in the opposite regime  $U = 1$, $V=2$ (red), while the dashed lines are the Poisson and the GOE limits.}
	\label{finite_size}
\end{figure}

Some of the most important foundations in the study of the relationship between quantum chaos and random matrices are the BGS and Berry conjectures, which are based on the common trend in the distribution of the nearest neighbors level spacing distribution between ensembles of Gaussian random matrices and sufficiently complex quantum systems. However, it is now clear in the literature that caution should be exercised when using this similarity as an indicator of the quantum chaos of the system, as several studies show that this distribution can also be found not because of the complexity of the system, but as a simple effect of finite size. It can be shown that even a non-chaotic system, such as a system described by the Anderson Hamiltonian, exhibits a transition in the distribution trend of the level spacing behavior from Poisson to Wigner-Dyson simply by varying the size of the system \cite{torres2019level,elkamshishy2021observation}. For this, we did a systematic analysis measuring the average level spacing ratio as a function of the size in both the \textit{near-integrable} and the opposite regime.\\
Fig.\ref{finite_size} shows, consistent with what has just been stated, that even in this system finite size effects can be observed in the energy level spacing statistics. However, the red curve shows that despite fluctuations, the value of the average level spacing ratio remains within the chaotic estimate. The blue line, on the other hand, normalizes its value around a more physical value than that at $S=5$, because the average level spacing ratio should tend from above to the Poisson limit, as a consequence of the small perturbation that cannot be integrated in $V$ from the integrable setting $U=1$ and $V = 1$.
This lends credibility to our study on the average level spacing ratio in Fig.\ref{UVplane}, since quantitatively the values may stabilize at slightly different numbers depending on U and V, but it is reasonable to assume that the system exhibits the same qualitative behavior in the parametric space and therefore the same physics.

\section{Fixed Points}
\label{appendix_fixed}
\begin{figure}[ht]
	\centering
	{\includegraphics[width=0.6\textwidth]{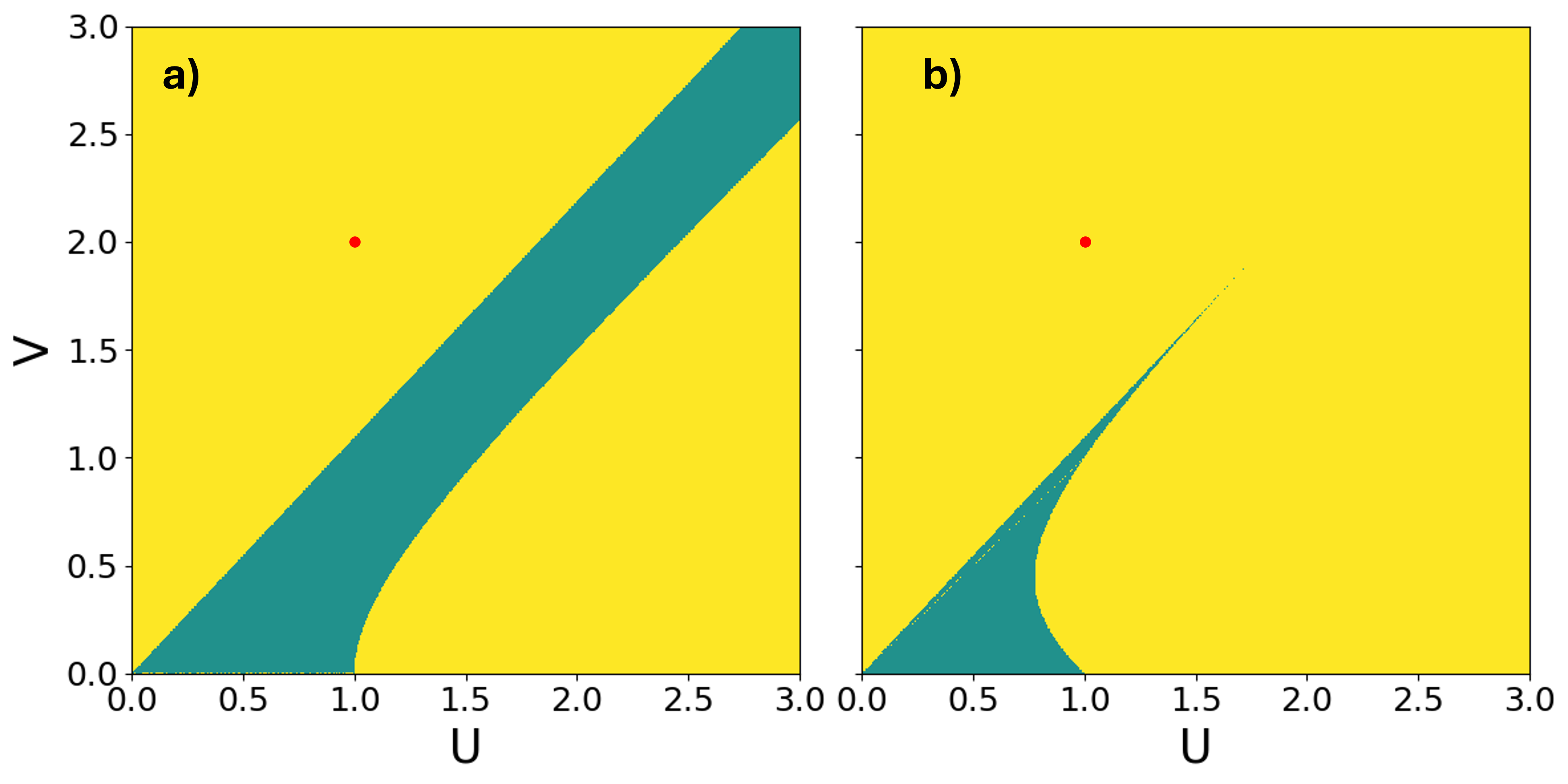}}
	\caption{Region of stability (sky blue) and the unstability (yellow) for (a) '$\pi00$-mode' and (b) '$\pi\pi0$-mode'. The scar analysis in the text is performed using the parameters corresponding to the red points.
	}
	\label{UVplanestab}
\end{figure}

Here we present the calculations relating to the stability of certain fixed points of the dynamic system in Eq. \eqref{dyn_sys_cl}. The most obvious ones derive from assigning each $z_i = 0$, where the only possible values for $\phi_i $ are 0 or $\pi$. The fixed points under analysis in this paper are those related to the '$\pi\pi0$ mode' (two $\phi_i = 0$) and the '$\pi00$ mode' (one $\phi_i = 0$). In order to study the stability of these fixed points, we perform a linear stability analysis, in which we neglect all second-order perturbations. In this way, the system can be easily reduced to a system involving only the variation variables on $z_i$:
\begin{equation}
	\label{pertpi00}
	\begin{cases*}
		\delta \ddot z_1 = - \delta z_1 - U \delta z_1 - V (\delta z_2 +  \delta z_{3})\\
		\delta \ddot z_2 = - \delta z_2 - U \delta z_2 - V (\delta z_1 +  \delta z_3)\\
		\delta \ddot z_3 = - \delta z_3 + U \delta z_3 - V (\delta z_1 +  \delta z_2)
	\end{cases*}
\end{equation}
for the '$\pi00$ mode'
\begin{equation}
	\label{pertpipi0}
	\begin{cases*}
		\delta \ddot z_1 = - \delta z_1 - U \delta z_1 - V (\delta z_2 +  \delta z_{3})\\
		\delta \ddot z_2 = - \delta z_2 + U \delta z_2 - V (\delta z_1 +  \delta z_3)\\
		\delta \ddot z_3 = - \delta z_3 + U \delta z_3 - V (\delta z_1 +  \delta z_2)
	\end{cases*}
\end{equation}
and for the '$\pi\pi0$-mode'.\\
Now, by initializing a harmonic perturbation $\delta z_i (t) = \delta z_i e^{i\omega t}$, the eigenfrequencies $\omega$ can be found through the characteristic equation of an eigenvalue problem of the matrices derived from the systems in eq.\eqref{pertpi00} and eq.\eqref{pertpipi0}, with eigenvalues $-\omega^2$:\\
\begin{equation}
	\begin{vmatrix}
		- (1+U) + \omega^2 & -V & -V\\
		- V & 	- (1+U) + \omega^2 & - V \\
		V & V & U-1+\omega^2
	\end{vmatrix}
\end{equation}
for the '$\pi00$-mode'
\begin{equation}
	\begin{vmatrix}
		- (1+U) + \omega^2 & -V & -V\\
		V & U-1+\omega^2 & V \\
		V & V & U-1+\omega^2
	\end{vmatrix}
\end{equation}
and for the '$\pi\pi0$-mode'.\\
Therefore, $\omega^2$ can be obtained by solving the cubic equation derived from the determinant of the appropriate matrix. Depending on the sign of $\omega^2$, the fixed point could be stable or unstable. Fig. \ref{UVplanestab} shows the stability region ($\omega^2 > 0$) colored in blue and the instability region ($\omega^2 < 0$) in yellow for both fixed points.
As we can see, the fixed points are unstable for a very large part of the $(U,V)$-parameter space (shown in yellow in the figure), where we can evaluate the presence of scarring phenomena on these states. As indicated by the red dot in parameter space in Figure \ref{UVplanestab}, $U = 1$ and $V = 2$ have been employed for the subsequent part on quantum system simulations.
\section{Time evolution of Husimi distribution}
\label{appendix_husimi}
\begin{figure}[ht]
	\centering
	\includegraphics[width=0.6\textwidth]{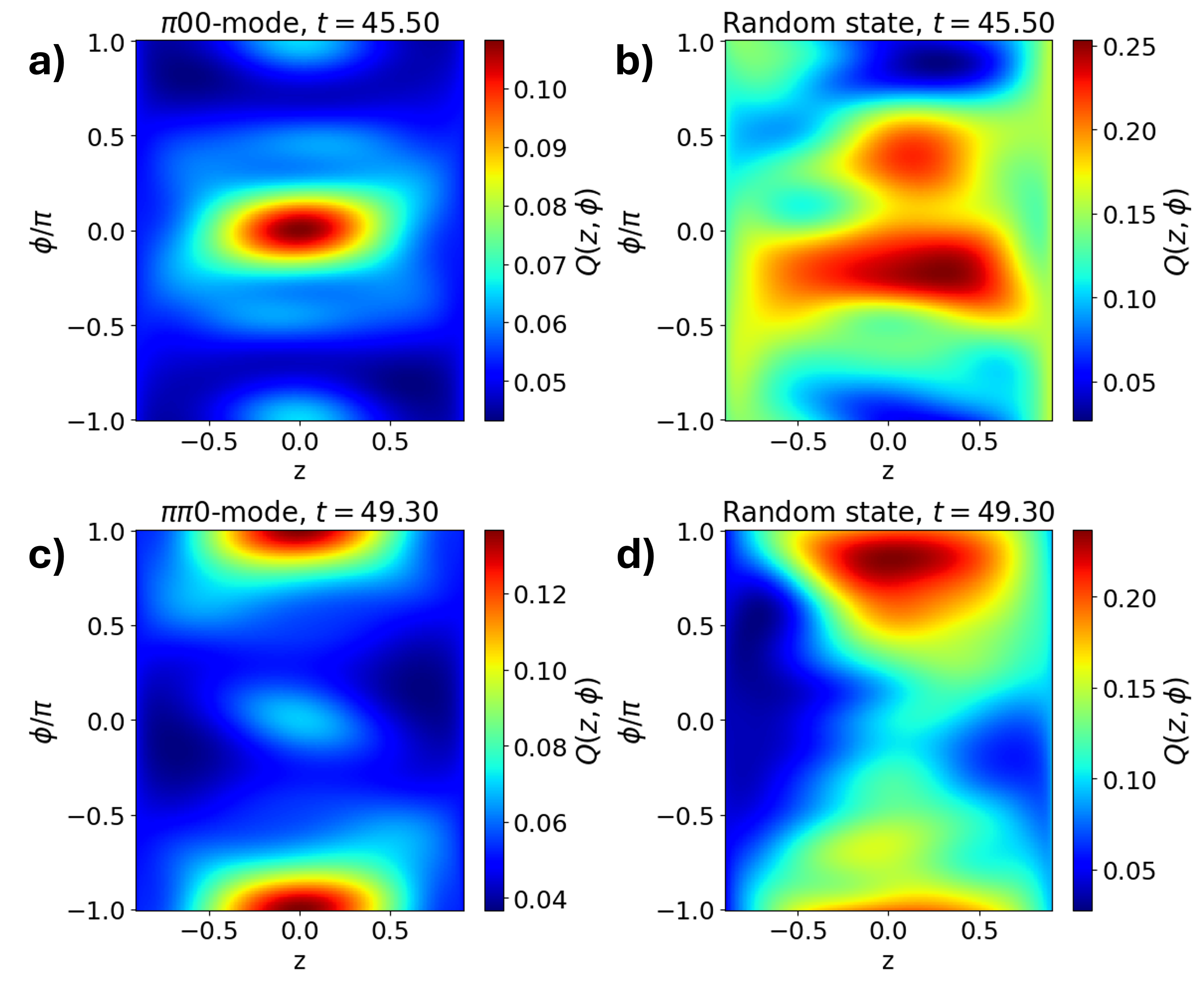}
	\caption{Panels (a) and (c) show the '$\pi00$' and '$\pi\pi0$' scar modes, respectively. For comparison, panels (b) and (d) display random coherent states at the same semiclassical energy of the corresponding scar. The snapshots are taken at times $t = 45.50$ for the '$\pi00$-mode' and $t = 49.30$ for the $\pi\pi0$-mode. The simulation was performed with each $S_i$ = 6 and parameters $U = 1, V = 2$.}
	\label{Husimi}
\end{figure}

These discussion about survival probability of quantum scars is further supported by a visualization of these states, provided by the Husimi distribution calculated over the variables of a single spin: 
\begin{equation}
	Q(z,\phi)=\frac{1}{\pi}\langle z,\phi|\hat\rho_1|z,\phi\rangle
\end{equation}
where the reduced density matrix on the first spin is built as
\begin{equation}
	\label{rhored}
	\hat\rho_1 = Tr_{23}(\hat\rho(0)) = Tr_{23}(|\psi(0)\rangle\langle\psi(0)|)
\end{equation}
and $|z, \phi\rangle$ is the coherent state of spin in the parameters $\tilde z$ and $\tilde\phi$ defined in the semiclassical system.\\
Figure \ref{Husimi}  shows the Husimi distribution at time $t=45.50\hspace{0.1cm}$ for the $\pi 0 0$-mode and its corresponding random coherent state and at $t=49.30\hspace{0.1cm}$ for the case of $\pi \pi 0$-mode. These two time are highlited by red points in the Figure \ref{Survival}, and are chosen as the time in which the scar state has the maximum overlap with itself at initial time.
The data confirm that states associated with $\pi$-modes behave as quantum scars, remaining coherent in the Husimi representation even at late times, retaining a memory of their initial configuration. This does not apply, however, to a state taken without a particular criterion at the same energy, such as the random coherent state, in which a diffusion of the packet in the $z-\phi$ space is evident.

\section{From separability of Hamiltonian to Thermalization}
\label{appendix_separability}
\begin{figure}[h]
	\centering	\includegraphics[width=0.6\textwidth]{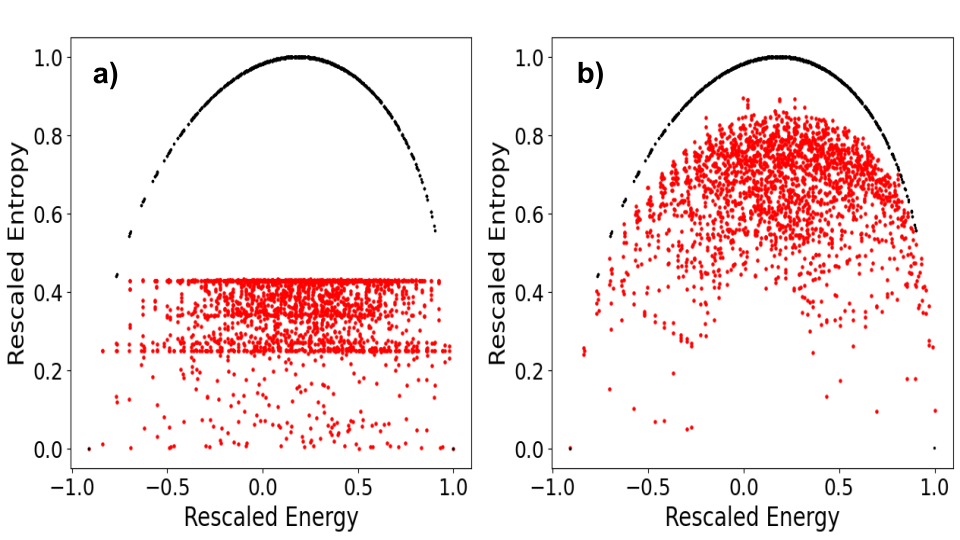}
	\hfill
	\caption{Rescaled entanglement entropy per single spin as a function of rescaled eigenenergy for $S = 6$. The theoretical thermal curve is derived from Eq. \eqref{ETH_rho}. Panel (a) shows a regime with a small deviation from separability ($V = 0.001$), while panel (b) shows a more significant deviation ($V = 0.1$).  The rescaled entropy indicates the entropy divided by the maximum entropy that can be obtained: $\ln (2S+1)$. The rescaled energy indicates the energy divided by the maximum eigenvalues of the system.}
	\label{EE_sep}
\end{figure}

The Hamiltonian presented in Eq. \eqref{H_BJJ_Schwin} exhibits a strictly separable structure when the interspecies coupling parameter $V$ vanishes. In this limit, the Hamiltonian decomposes into the sum of independent terms:
\begin{equation}
H = \sum_{\alpha = 1,2,3} H_{BJJ,\alpha},
\end{equation}
where each $H_{BJJ,\alpha}$ acts as a local constant of motion. Because this structure renders the total Hamiltonian strictly separable, the resulting eigenstates are product states within the single-spin basis. Consequently, these states fail to satisfy the Eigenstate Thermalization Hypothesis, as the system lacks the requisite entanglement between subsystems. As shown in Fig. \ref{EE_sep}, increasing the parameter $V$ breaks this separable structure, introducing non-trivial coupling and entanglement between the bosonic species. This departure from separability acts as the underlying mechanism driving the transition toward thermal behavior.

\section{Integrable Limit}
\label{appendix_integrable}
The Hamiltonian in Eq.\eqref{H_BJJ_Schwin} can be reduced into simpler form in the special case in which $U=V$. As a matter of fact, in this case the Hamiltonian becomes:
\begin{equation}
	H = - \sum_{\alpha = 1,2,3} S_{\alpha x}+ \frac{U}{2S} \left(\sum_{\alpha = 1,2,3}  S_{\alpha z}\right)^2
\end{equation}
In the following we prove that there emerge some other non trivial constants of motion in this case that make the system integrable. The new independent integral of motions are chosen to be $(\vec S_1+\vec S_2+\vec S_2)^2$ and  $(\vec S_1+\vec S_2)^2$.\\
Let's introduce the composed angular momentum $\left(\sum_{\alpha = 1,2,3}  \vec S_{\alpha}\right)^2 = S_{123}^2$. We can note that in the special case in which $U=V$, the operator $S_{123}^2$ commutes with Hamiltonian because $[S_{123}^2,S_{123,j}]=0$ for every \textit{j}. This is the first emergent constant of motion.\\
Now we analyze the composed angular momentum $(\vec S_1+\vec S_2)^2 = S_{12}^2$. We can rewrite the Hamiltonian as 
\begin{equation}
	\begin{split}
		H &= - S_{12,x} - S_{3x} + \frac{U}{2S} \left(S_{12,z} + S_{3z}\right)^2\\
		&=  - S_{12,x} - S_{3x} + \frac{U}{2S} (S_{12,z}^2 + S_{3z}^2 + 2S_{12,z} S_{3z})
	\end{split}
\end{equation}
Since every addendum of the sum commutes with the operator $S_{12}^2$, and thus it represents a constant of motion.\\
The last thing that remains to be checked to verify the integrability of the problem is that $[S_{123}^2,S_{12}^2]=0$.\\
It can be easily seen that
\begin{equation}
	[S_{123}^2,S_{12}^2]=4[\vec{S_2}\cdot\vec{S_3},\vec{S_1}\cdot\vec{S_2}] + 4[\vec{S_3}\cdot\vec{S_1},\vec{S_1}\cdot\vec{S_2}]
\end{equation}
Then writing explicitly $\vec{S_i}\cdot\vec{S_j} = S_{ix}S_{jx} + S_{iy}S_{jy} + S_{iz}S_{jz}$, and by exploiting the commutation rules between angular momenta it can be proved that $[\vec{S_3}\cdot\vec{S_1},\vec{S_1}\cdot\vec{S_2}] = -[\vec{S_2}\cdot\vec{S_3},\vec{S_1}\cdot\vec{S_2}]$ and thus $[S_{123}^2,S_{12}^2]=0$. \\
This final check demonstrates the integrability of the system.

\section{Scaling of Entanglement Entropy with System Size}
\label{appendix_EE_scaling}
\begin{figure}[h]
	\centering	\includegraphics[width=0.5\textwidth]{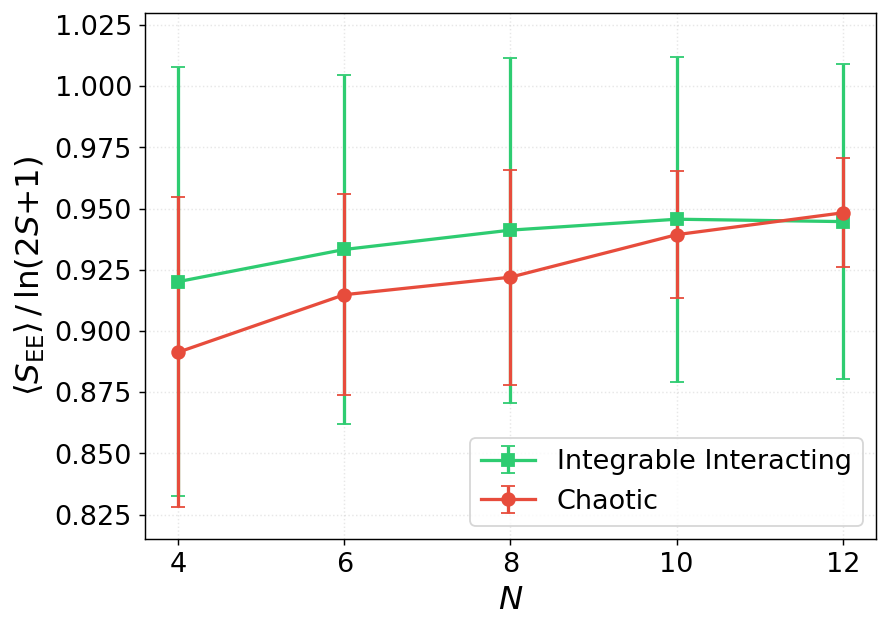}
	\hfill
	\caption{Average EE in the middle of the spectrum as a function of number of particles $N = 2S$.  The error bars are representing $\sqrt{Var(EE)}/\ln(2S+1)$.}
	\label{EE_size}
\end{figure}

The entanglement entropy (EE) of symmetry resolved eigenstates exhibits significant fluctuations, reflecting the fact that the Hilbert space dimension follows a power law: $D = (2S+1)^3 = (N + 1)^3 \sim N^3$. To substantiate our qualitative findings, we analyze the dependence of the EE in the middle of the spectrum as a function of the system size $N = 2S$ for both the integrable interacting and non-integrable regimes. Fig. \ref{EE_size} represents the average of the EE for eigenstates near the center of the spectrum as a function of size showing a clear volume-law behavior. 

\section{Entanglement Entropy for Degenerate Eigenstates}
\label{appendix_EE_sectors}
\begin{figure}[ht]
	\centering	\includegraphics[width=0.9\textwidth]{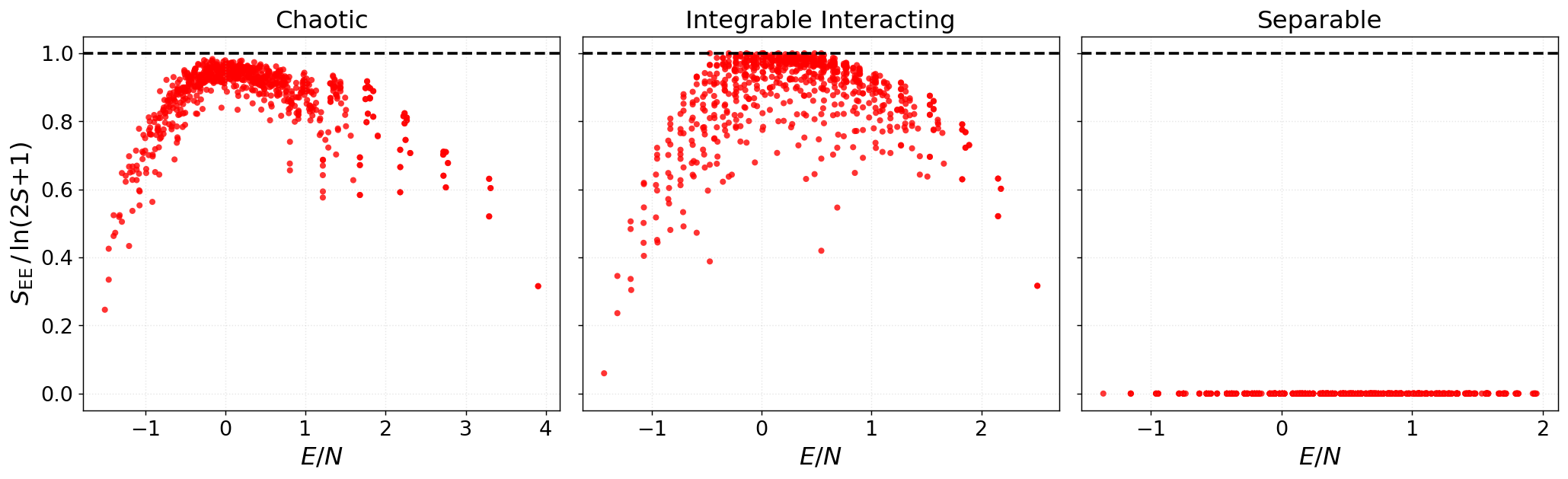}
	\hfill
	\caption{The rescaled Entanglement Entropy as a function of the rescaled eigenenergy for each eigenstate. From left to right: the non-integrable regime ($V=2U=2$), the integrable interacting regime ($U=V=1$), and the separable regime ($V=0$).  The horizontal black dashed line is referred to the maximum possible entropy $\ln(2S+1)$. Here $N = 2 S = 8 $, $d = 2S + 1$.}
	\label{EE_sectors}
\end{figure}

In the main text, the entanglement entropy presented in Fig. \ref{EE} is computed for degenerate eigenstates without resolving symmetry sectors. While this approach is sufficient to capture the general behavior, the specific values may depend on the superposition of degenerate eigenstates. To demonstrate the robustness of our results, we show in Fig. \ref{EE_sectors} that we obtain the same qualitative conclusions when the eigenstates are selected from the specific symmetry sectors of the Hamiltonian (see Appendix \ref{appendix_symmetry}). Both the non-integrable and the interacting integrable regimes continue to exhibit volume-law scaling. In the separable case, the entropy remains obviosuly zero when resolved within the appropriate symmetry sectors.

\end{appendix}
\bibliography{Referenze}
\end{document}